\documentstyle[11pt]{article}
\begin{document}

\title{Miscellaneous Stars in the OGLE Catalog of Periodic Variable
Stars in the Galactic Bulge}

\author{A~r~k~a~d~i~u~s~z ~~ O~l~e~c~h}

\date{Warsaw University Observatory, Al.~Ujazdowskie~4,~00-478~Warszawa,
Poland}
\maketitle

\abstract{The first catalog of chromospherically active and ellipsoidal
variable stars in the OGLE database is presented. The periods, the observed 
magnitudes and colors, their values free of interstellar extinction, and
Fourier coefficients of light modulation are 
given for 549 stars. New color--magnitude diagrams based on new color
determinations are presented.  The chromospherically active stars are
all among the bulge red giants, while the ellipsoidal binary stars are
near the disk main sequence turn--off point.
A discovery of a small group of very red ($2<(V-I)_0<6$) and bright,
chromospherically active bulge giants ($I_0\approx13$ mag) is also reported.}

\section{Introduction}

The Optical Gravitational Lensing Experiment (OGLE) is a long term
observing project of the Warsaw University Observatory, the
Princeton University Observatory and the Carnegie Institution of Washington,
which begun in 1992.  Its main goal is a search for dark matter in our
Galaxy using the microlensing phenomena (Paczy\'nski 1986, Udalski {\it et al}
1992). For this purpose the CCD photometry of a few millions stars in dense
Baade's Window in Galactic bulge was performed.  All the data was
obtained with the 1-m Swope telescope at the Las Campanas Observatory in Chile,
operated by Carnegie Institution of Washington, with a $2048\times2048$ 
Ford/Loral CCD detector.  In four seasons of photometric observations 
(1992--95) nineteen microlensing events have been detected (Udalski
{\it et al} 1993, Udalski {\it et al} 1994a, Udalski {\it et al} 1994b, 
Paczy\'nski \& Udalski 1996).

Such a large amount of precise photometric data gives us opportunity to
obtain many light curves of variable stars in the Galactic bulge. The first
results of the OGLE search for periodic variables in Baade's Window were
published by Udalski {\it et al} (1994c, 1995a, 1995b, 1996) as The Catalog of
Periodic Variable Stars in the Galactic Bulge. The first three parts of the
catalog presented results for the central part of Baade's Window (nine
$15'\times15'$ fields centered at $\alpha=18^h03^m24^s ~
\delta=-30^\circ02'00''$), the fourth part contains data for three
$15'\times15'$ fields on the eastern side of Baade's Window. 
The Catalog contained stars with $\langle I \rangle$ in the
range 14--18 mag. The upper limit $I\approx14$~mag is a result of
saturation of stellar images on CCD frames, and the lower one is
due to rapidly increasing photometric errors. 
Although the period search was
limited to periods from 0.1 day to 100 days, there are a few $\delta$
Scuti stars with periods $<0.1^d$ which were identified with $2 \times P$,
and a few Mira type variables with periods $>100^d$.

The variable stars discovered in every field were grouped into three
categories: pulsating stars (mostly RR Lyr and $\delta$ Sct stars),
eclipsing stars (mostly W UMa, $\beta$ Lyr and Algol type) and
miscellaneous variables (mostly late type, chromospherically active
stars, some ellipsoidal variables and few Miras).

Equatorial coordinates (epoch 2000.0) of the BW fields, the number of
the variable stars detected in each category are given in Table 1.
\vspace{0.5cm}
\begin{center}
Table 1 \\
\vspace{0.25cm}
The equatorial coordinates and numbers of detected stars in the OGLE Catalog.
\\
\vspace{0.4cm}
\begin{tabular}{|c|c|c|c|c|}
\hline
{BW} & {Equatorial} & {Pulsating} & {Eclipsing} & {Miscellan.} \\
{Field} & {Coordinates} & {Stars} & {Stars} & {Stars}
\\
\hline
BWC & $18^h03^m24^s ~~ -30^\circ02'00''$ & 31 & 116 & 66 \\
BW1 & $18^h02^m24^s ~~ -29^\circ49'05''$ & 24 & 128 & 70 \\
BW2 & $18^h02^m24^s ~~ -30^\circ15'05''$ & 13 & 100 & 56 \\
BW3 & $18^h04^m24^s ~~ -30^\circ15'05''$ & 18 & 113 & 74 \\
BW4 & $18^h04^m24^s ~~ -29^\circ49'05''$ & 16 & 124 & 64 \\
BW5 & $18^h02^m24^s ~~ -30^\circ02'05''$ & 18 & ~91 & 65 \\
BW6 & $18^h03^m24^s ~~ -30^\circ15'05''$ & 16 & ~83 & 54 \\
BW7 & $18^h04^m24^s ~~ -30^\circ02'05''$ & 18 & ~93 & 55 \\
BW8 & $18^h03^m24^s ~~ -29^\circ49'05''$ & 12 & ~85 & 54 \\
BW9 & $18^h00^m50^s ~~ -29^\circ49'05''$ & 19 & 125 & 81 \\
BW10 & $18^h00^m50^s ~~ -30^\circ02'05''$ & 23 & 130 & 76 \\
BW11 & $18^h00^m50^s ~~ -30^\circ15'05''$ & 17 & ~93 & 67 \\
\hline
Tot. & --- & 225 & 1281 & 782 \\
\hline
\end{tabular}
\end{center}
\vspace{0.6cm}

The main aim of this paper is to provide more information for the 
``miscellaneous'' variable stars in the BWC--BW8 fields. In our work we
omitted the stars unambiguously classified as Mira type variables, and
we obtained 549 stars suspected of chromospheric activity or
ellipsoidal variability. The additional 224 miscellaneous stars from fields
BW9--BW11 are not analyzed as there is no extinction map for
this part of the sky.

\section{Light Curves.}

The OGLE Catalog of Periodic Variable Stars in Galactic Bulge contains
an atlas with phased light curves and $30'' \times 30''$ finding charts.
The photometric data presented in the Catalog is available to
astronomical community in electronic form via INTERNET (ftp host: 
{\it sirius.astrouw.edu.pl}, directory {\it /ogle/var\_catalog}). The
photometry of each star is based on four seasons of observations (from
1992 to 1995). It is clearly visible that many stars classified as
miscellaneous change their light curves from season to season. 
The most prominent example of such behavior is provided by the
variable BWC~V34. The $I$ and $V$ photometry of this
star is presented on Fig. 1. On the other hand, there are variables with
low amplitude, almost sinusoidal light curves, which do not vary
from season to season.  The variable BWC~V39 is representative of this 
group.  The $I$ and $V$ light curves of this star are shown on
Fig. 2. It is not hard to notice that the first group contains
chromospherically active stars with spotted surface and the second one
contains ellipsoidal variables. Unfortunately, distinguishing between these
groups is not always as obvious as in the examples shown in Figs. 1 and 2.

It is desirable to be able to classify various light curves according
to some algorithm, rather than by visual inspection.  With a hope
to make it possible we decided to construct Fourier fit to the
light curves in the form:

\begin{equation}
I=a_0 + a_1 \cdot \sin 2\pi x + a_2 \cdot \sin 4 \pi x + b_1 \cdot \cos 2
\pi x + b_2 \cdot \cos 4 \pi x\end{equation}

\noindent The fitting was done for every light curve in every season.
The coefficients $a_0, a_1, a_2, b_1, b_2$ were calculated by the Least
Squares Method. Variable $x$ is defined as $ x \equiv {{t-t_0}\over{P}}$
and $t_0$ is the time of the first observation.
Table 2 summarizes our results. The columns of this
Table contain the following informations:
\vspace{0.2cm}

\noindent 1. ~Star designation. The letters {\it a}, {\it b}, {\it c},
{\it d} correspond respectively to seasons 1992, 1993, 1994, 1995.

\noindent 2. ~Period in days.

\noindent 3. ~Value of $a_0$.

\noindent 4. ~Value of $a_1$.
 
\noindent 5. ~Value of $b_1$.
 
\noindent 6. ~Value of $a_2$.

\noindent 7. ~Value of $b_2$.
\vspace{0.2cm}

The full version of Table 2 is available in electronic form via INTERNET
from URL {\it http://www.astrouw.edu.pl/$\sim$olech/misc.html}.

Fig. 3 and Fig. 4 give $I$ light curves for all seasons for the
above-mentioned representatives of our groups. Solid line
corresponds to the fit given by the equation (1). Additionally, the values of
$a_1$ and $b_1$ for twenty randomly chosen stars are shown on Fig. 5.

\section{New Color Determination.}

The first edition of the Catalog published by Udalski {\it et al} (1994c,
1995a, 1995b, 1996) gave $V-I$ colors of variable stars based
on a very few $V$-band measurements made during the first three seasons.
The accuracy of $V$ magnitude was $ \sim 0.05 $ mag.
The $V-I$ color was estimated only at maximum brightness.
In the 1995 season many new $V$-band observations were made, and
in many cases it became possible to obtain good $V$ light curves.
Therefore, we decided to make new color determinations.
The first step of this procedure was to fit the Fourier function (1) to
the phased $I$ light curves for the fourth season. 
Next, for the phase corresponding to every $V$ point we 
calculated the $I$ magnitude with the fitted
function. The final $V-I$ color of the star was obtained as an average
value of all such determinations.

The new color--magnitude diagram (CMD) presented in Fig. 6 is based on
new color determinations. The open circles represent miscellaneous
variable stars and dots stars with constant brightness. Only 20\% of non
variable stars are plotted for clarity.

Recently Wo{\'z}niak \& Stanek (1996) proposed a new method of
investigating interstellar extinction, based on the two band ($V$ and $I$)
photometry, which uses the red clump stars to construct the reddening
curve. Stanek (1996) applied this method to CMDs obtained by the OGLE
collaboration and he constructed the extinction map of the central part of
Baade's Window (fields BWC--BW8). We used this reddening map to calculate 
values of $I_0$ and $(V-I)_0$ free from extinction.  Our results are summarized
in Table 3. This table gives the star designation, its period in days, $I$,
$V-I$, $I_0$, $(V-I)_0$, the mean amplitude of the brightness of
the star and the variation of this amplitude. The values of $I_0$ and
$(V-I)_0$ are used to plot true CMD presented on Fig. 7. In this Figure
only 5\% of the non variable stars were plotted.

It is worthwhile to note that the values of interstellar extinction
given by Stanek (1996) are correct only for the Galactic bulge stars,
and not in the disk stars. Therefore the main sequence on our CMDs might be
bluer than it reality should be.

\section{Distinguishing Between Chromospherically Active and
Ellipsoidal Variable Stars.}

We attempted to divide the miscellaneous stars into two
groups; the first one containing chromospherically active stars, and the
second one with ellipsoidal variables. We expected that
chromospherically active stars, due to their changing light curves, should
have a large scatter of their amplitudes. Therefore we plotted a
correlation between the mean amplitude of the modulation of each star
and Gaussian variation -- $\sigma$ of this amplitude from season to season. The
result of this operation is shown in Fig. 8. Unfortunately, there is no
clear distinction between the two groups, perhaps because
many chromospherically active stars have small amplitudes.

Next, we tried another approach to separate the two types of variables.
If a variable with a small amplitude and a sinusoidal light curve is an
ellipsoidal variables then its true period should be twice longer than the
one listed in
Table 2 and 3. Therefore, the light curve of such a star should be better
described by coefficients $a_{0.5}$ and $b_{0.5}$ (corresponding to the 
period $2\times P$) than $a_1$ and $b_1$. This
property could be visible in a graph showing the dependence of
$\sqrt{a_1^2+b_1^2}$ on $\sqrt{a_{0.5}^2+b_{0.5}^2}$. Such a graph is
presented in Fig. 9.  Unfortunately, no clear separation into two groups is
apparent.

The next step was to plot the period of a star versus 
its brightness, as shown in Fig 10a. All limits of the Catalog
are clearly visible. The upper limit near $I=14$~mag is a result of 
a saturation of stellar images on the CCD frames. The $I=18$~mag is
a limiting magnitude of the present Catalog. The left and
right boundaries result from the range of period search. A very
interesting feature is the absence of stars in the upper--left part of the
Figure. In order to check its physical nature we slightly modified our
graph. We changed vertical axis from $I$ to $I_0$, and the absence of stars
in the upper--left part of Fig. 10b is just as clear as it was in Fig. 10a.
This phenomenon is not due to the Catalog, but it has a physical cause: a 
variable of a given luminosity cannot have a period shorter than some limit.
The solid line on Fig. 10b represents the minimal period of a rotating 
single star rotating at the so called ``break-up''.
The line was obtained
in the following way: we adopted a linear relation between $V_0$ and
$(V-I)_0$, as calculated with the Least Squares Method:
$(V-I)_0=0.07574\cdot V_0+0.03272$. The dependence between $(V-I)_0$ and
the effective temperature $T_{eff}$ of a star was taken from Bartelli
{\it et al} (1994), with $T_{eff}=5000~K$ corresponding to
$(V-I)_0=0.97$~mag and  $T_{eff}=3500~K$ to $(V-I)_0=2.40$~mag. Assuming
a typical stellar mass as 1$\cal M_\odot$ and the distance of 8~kpc we
obtained a relation which may be well approximated with:

\begin{equation}
I_0=-3.58\cdot \log P + 16.05
\end{equation}
where the period $P$ is measured in days.

The line in Fig. 10b is a good envelope to the distribution of stars,
and this demonstrates that the absence of stars in the upper--left part
of the Figure is due to the physical limit.
Only one star is placed considerably
above the limiting line for a single rotating star. It might be a
one star which is a result of mixing two stars of a binary system.

The color--magnitude diagram presented in Fig. 7 contains 
a few blue stars with $(V-I)_0<1$,
which are placed mainly at the Galactic disk main sequence turn-off point.
These stars are clearly
separated from other miscellaneous variables, which are distributed
above the bulge main sequence turn off point, on the red sub giant and
the red giant branch. We decided to use a subjective judgement about
the nature of these blue variables, and for every star with
$(V-I)_0<1$ we classified its light curve: those with most regular
variations were expected to be ellipsoidal variables, those with the
most irregular light curves were expected to be chromospherically
active.  Three groups were formed, and they are listed in Tables 4a, 4b, 4c.

Table 4 contains
periods, magnitudes and colors of those blue stars which have light curves
strongly resembling ellipsoidal variables. It is worthwhile to
note that almost all stars so selected have very short periods and
colors in the range $0.3<(V-I)_0<0.8$. Table 5 contains blue stars which
light curves could belong to ellipsoidal variables but its connection
with chromospherically active stars cannot be excluded. It is clearly
visible that periods of these stars are not restricted to such a small
range as in Table 4, but in many cases the periods are still short. The
colors of these stars are mostly in the range 0.8--0.9 mag. The last
group is presented in Table 6 and contains blue stars which light curve
definitely excluded connection with the ellipsoidal variability. The range
of periods of stars in this group is very large, and the colors
are very close to 1.0 mag.

We believe that objects listed in Table 4
and the majority of these listed in Table 5 are ellipsoidal
variable stars. These blue stars are shown in the color--period
diagram in Fig. 11 with open circles, and they form
a narrow branch of a few points with blue color and short period. 

The full list of OGLE stars with V and I data contains 528~138 objects.
It is not a precise estimate because of photometric errors (see for
comparison Szyma\'nski {\it et al} 1996). We estimate that in area the
below red clump i.e. for $14.5<I_0<16$ and $(V-I)_0>1$,
there are 24~744 constant stars. In the same region there are 273
miscellaneous variable stars. It means that about 1\% of all stars in 
that part of CMD
is chromospherically active with amplitude of modulation large enough to
be detected in the OGLE search. Of course there are variables classified
as miscellaneous also outside the above-mentioned area, but 
their highest concentration is observed inside this area.

Two normalized histograms of color distribution in 0.1 mag bins
are presented in Fig. 12 for stars with $15.0<I_0<16.0$. The solid and dotted 
lines correspond to constant and to variable
stars respectively. The distribution of the colors in both cases seems
to be similar, but the dotted line is slightly redder than the solid one.

\section{Red stars.}

Among the few hundred miscellaneous variables
we found eight exceptionally red stars which are clearly visible
in the CMD presented in Fig. 13. Our first impression was that $V$-band
brightness of these star was measured incorrectly. Therefore, we
investigated $V$-band light curves of these stars and we found that
they are phased with the same periods as $I$-band light curves. A large
number of $V$ data points excludes a possibility of a
mistake, as demonstrated in Fig. 14 and 15,
where $I$ and $V$ light curves of these stars are presented.

The following stars belong to this group are BW1~V1, BW3~V1, BW3~V2,
BW5~V5, BW6~V3, BW7~V1, BW7~V5, BW8~V2 and also variable V1 in the field
BW9. These red stars are distinguished not only by the large values of 
$(V-I)_0$ but also by their high brightness and long periods. 
They are often the brightest variable stars
in their fields. It would be interesting to search for 
such red variables among the stars brighter than $I\approx14$, which is
the upper limiting magnitude of the Catalog.

Previous surveys of Baade's Window region (Frogel \& Whitford 1987,
Blanco {\it et al} 1984, Lloyd Evans 1976) also revealed examples of such red
and bright objects. The majority of the reddest stars in those surveys was
variable, and because of their large amplitudes
and long periods they were classified as Mira
type variables. Additionally a large part of their variables was
semiregular. The red stars presented in this paper are certainly not
Miras, but they may be semiregular variables.
Their periods are shorter than 100
days, their light curves are unstable and change from season to season. It
suggests that these objects belong to the group of chromospherically
active variables, but we cannot exclude the possibility that they
are semiregular variables. 

Similar red and bright objects were also found in a few open clusters. For
instance, Garnavich {\it et al} (1994) reported observations of the red giant
branch in old and metal rich open cluster NGC 6791. They found at least
12 such stars. These objects are placed at the end of evolutionary sequences
of stars with masses near 1.1$\cal M_\odot$. In the same
cluster Ka{\l}u\.zny \& Ruci\'nski (1993) discovered 17 variable stars.
One of them was very red $B-V=1.595$ evolved giant and was located above
the horizontal branch red clump. They suggested that variability of this
star was caused by pulsations and classified it as RV Tau--type star.

\section{Conclusion and Summary}

We have presented the first catalog of stars classified as
miscellaneous in the OGLE search for variable stars in the Galactic bulge. Our
catalog contains 549 objects. The periods, the magnitudes and colors as
observed and free of interstellar extinction, the 
coefficients of Fourier fit and
amplitudes of the light modulation are given for each star. 517 object
are suspected for chromospherical activity. 
Some of these stars are likely to
be single rotating spotted stars of the FK Com type.
Other objects from this group could be binary systems with
spotted component, like RS CVn binaries. 32 stars with colors
$(V-I)_0<1$, short periods and sinusoidal, small amplitude light curves
are classified as ellipsoidal variables i.e. binary systems with
ellipsoidal components not showing eclipses.

A group of 8 stars from our list is distinguished by the very large
values of $(V-I)_0$ color (in range 2.5--6 mag), and high and similar
$I$--band brightness, near the bright limit of the OGLE catalog. 
These are likely to be chromospherically active
red giants. It is interesting that there are no
constant stars at the part of CMD where our reddest variables are
located. Similar property was presented by Frogel \& Whitford (1987)
where variables were also the reddest M giants found. It might suggests
that the spots on the stellar star cause a considerable reddening of
the star.

The chromospherically active stars in the are almost uniformly spread along
red subgiants, giant and super giant branches, where they constitute about 1\%
of all stars in the OGLE database.  There is no clear concentration of these 
variables in the red clump region of the CMD,
which suggest that spotted stars evolve for the first time
along the red giants branch.
\vspace{0.1cm}

\noindent {\bf Acknowledgments} ~I am truly grateful to Prof. Bohdan 
Paczy\'nski for his
helpful discussion, reading and commenting on the manuscript.

\begin{center}
REFERENCES
\end{center}
\vspace{0.2cm}

\vspace{5pt}
\noindent{Bartelli, G., Bressan, A., Chiosi, C., Fagotto, F., and Nasi,
E.}~{1994}~{\it Astron. Astrophys. Suppl. Ser.}~{106}~{275}

\vspace{5pt}
\noindent{Blanco, V.M., McCarthy, M.F., and Blanco, B.M.}~{1984}~{\it Astron.
J.}~{89}~{636}

\vspace{5pt}
\noindent{Frogel, J.A., and Whitford, A.E.}~{1987}~{\it ApJ}~{320}~{199}

\vspace{5pt}
\noindent{Garnavich, P.M., VandenBerg, D.A., Zurek, D.R., and Hesser,
J.E.}~{1994}~{\it Astron. J.}~{107}~{1097}

\vspace{5pt}
\noindent{Ka{\l}u\.zny, J, and Ruci\'nski, W.}~{1993}~{\it MNRAS}~{265}~{34}

\vspace{5pt}
\noindent{Lloyd Evans, T.}~{1976}~{\it MNRAS}~{174}~{169}

\vspace{5pt}
\noindent{Paczy\'nski, B.}~{1986}~{\it ApJ}~{304}~{1}

\vspace{5pt}
\noindent{Paczy\'nski, B., and Udalski, A.}~{1996}~{\it The proceedings of
12th IAP Colloquium: Variable Stars and the
Astrophysical Returns of Microlensing Surveys}

\vspace{5pt}
\noindent{Stanek, K.Z.}~{1996}~{\it ApJ Letters}~{460}~{37}

\vspace{5pt}
\noindent{Szyma\'nski, M., Udalski, A., Kubiak, M., Ka{\l}u\.zny, J.,
Mateo, M., and Krzemi\'nski, W.}~{1996}~{\it Acta Astron.}~{46}~{1}

\vspace{5pt}
\noindent{Udalski, A., Szyma\'nski, M., Ka{\l}u\.zny, J., Kubiak, M., and 
Mateo, M.}~{1992}~{\it Acta Astron.}~{42}~{253}

\vspace{5pt}
\noindent{Udalski, A., Szyma\'nski, M., Ka{\l}u\.zny, J., Kubiak, M.,
Krzemi\'nski, W., Mateo, M., Preston, G.W., and Paczy\'nski,
B.}~{1993}~{\it Acta Astron.}~{43}~{289}

\vspace{5pt}
\noindent{Udalski, A., Szyma\'nski, M., Stanek, K.Z., Ka{\l}u\.zny, J.,
Kubiak, M., Mateo, M., Krzemi\'nski, W., Paczy\'nski, B., and Venkat,
R.}~{1994a}~{\it Acta Astron.}~{44}~{165}

\vspace{5pt}
\noindent{Udalski, A., Szyma\'nski, M., Ka{\l}u\.zny, J., Kubiak, M.,
Mateo, M., Krzemi\'nski, W., and Paczy\'nski, B.}~{1994b}~{Acta
Astron.}~{44}~{227}

\vspace{5pt}
\noindent{Udalski, A., Kubiak, M., Szyma\'nski, M., Ka{\l}u\.zny, J.,
Mateo, M., and Krzemi\'nski, W.}~{1994c}~{\it Acta Astron.}~{44}~{317}

\vspace{5pt}
\noindent{Udalski, A., Szyma\'nski, M., Ka{\l}u\.zny, J., Kubiak, M.,
Mateo, M., and Krzemi\'nski, W.}~{1995a}~{\it Acta Astron.}~{45}~{1}

\vspace{5pt}
\noindent{Udalski, A., Olech, A., Szyma\'nski, M., Ka{\l}u\.zny, J., Kubiak, M.,
Mateo, M., and Krzemi\'nski, W.}~{1995b}~{\it Acta Astron.}~{45}~{433}

\vspace{5pt}
\noindent{Udalski, A., Olech, A., Szyma\'nski, M., Ka{\l}u\.zny, J.,
Kubiak, M., Mateo, M., Krzemi\'nski, W., and Stanek K.Z.}~{1996}~{Acta
Astron.}~{46}~{51}

\vspace{5pt}
\noindent{Wo\'zniak, P., and Stanek, K.Z.}~{1996}~{\it ApJ}~{464}~{233}

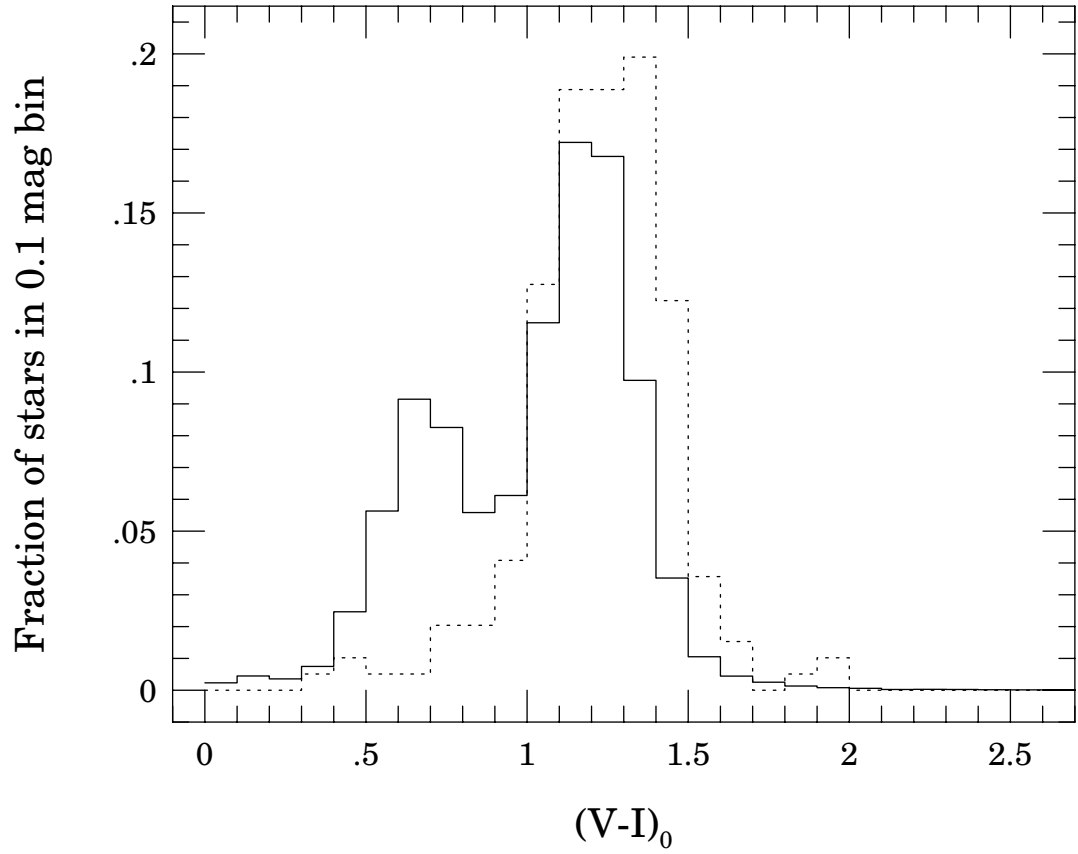
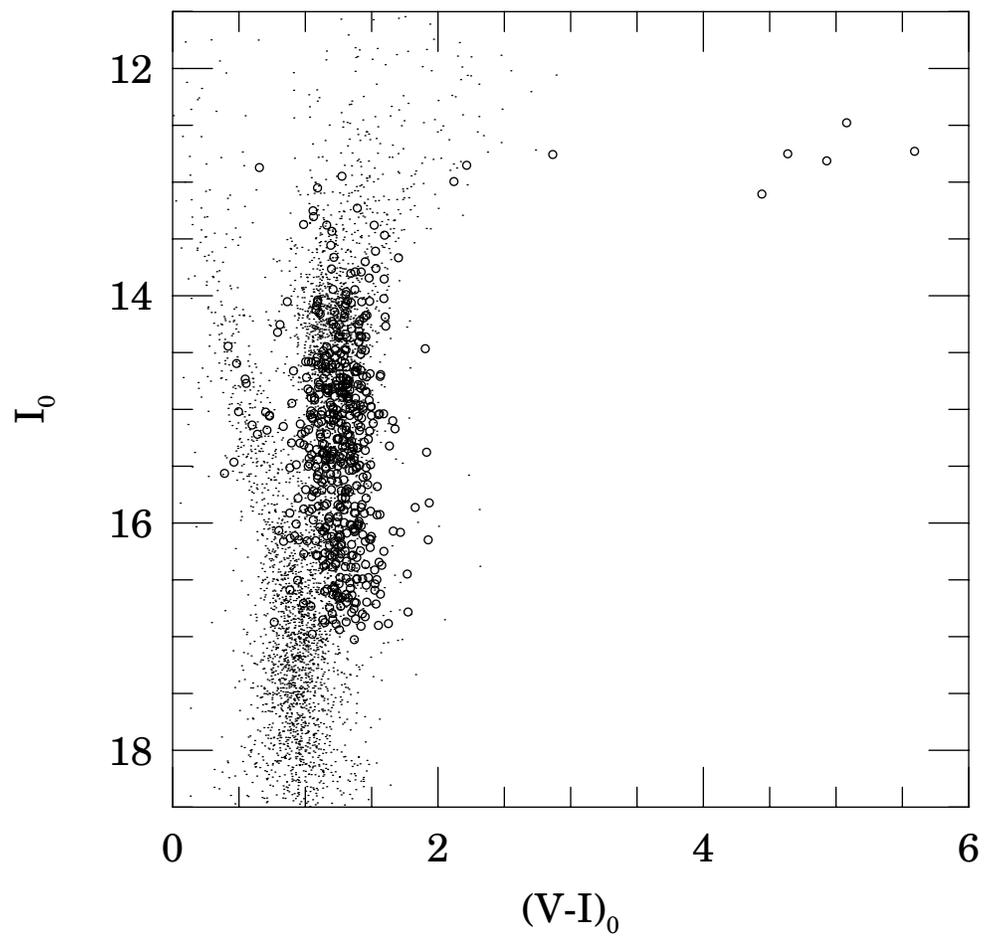
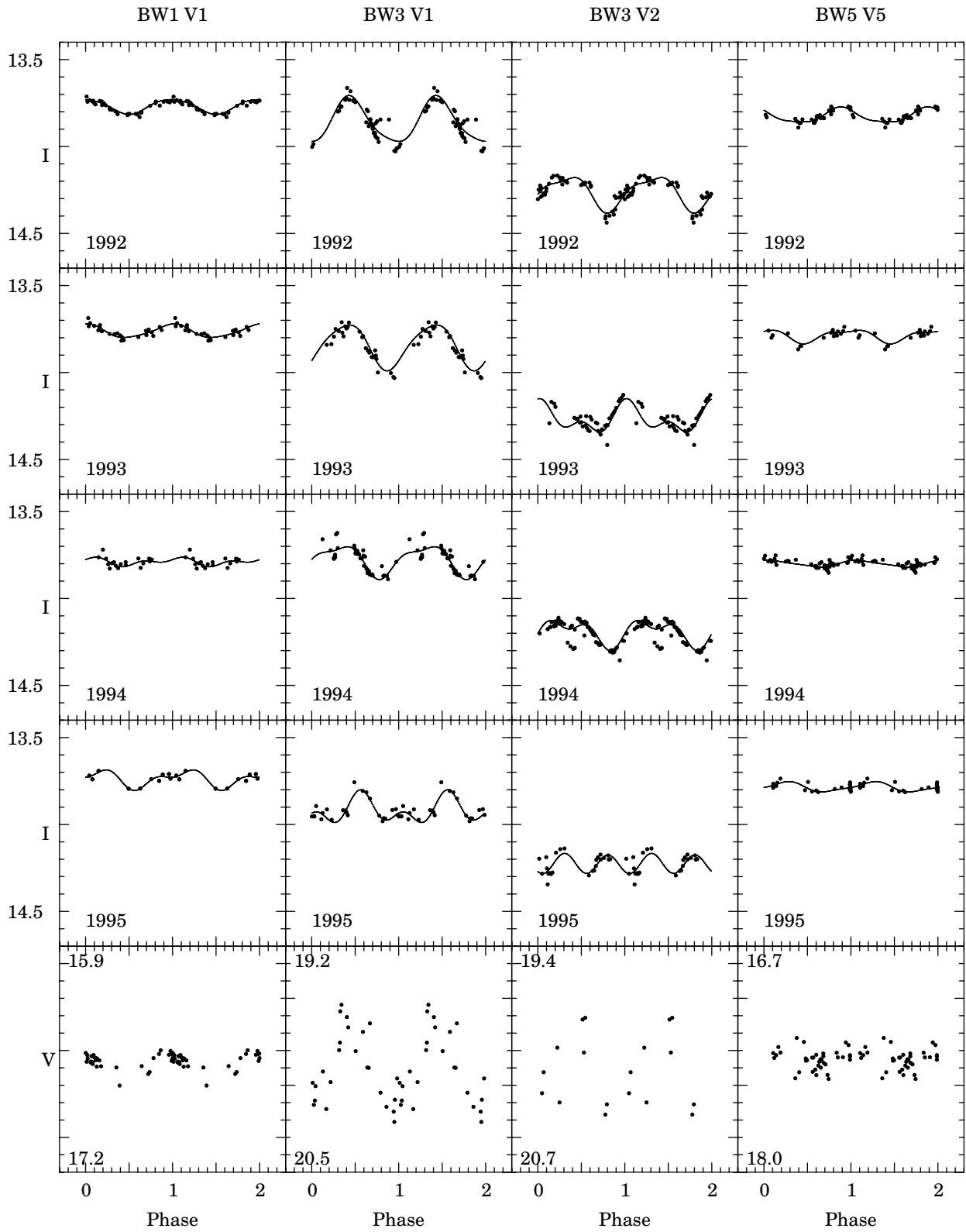
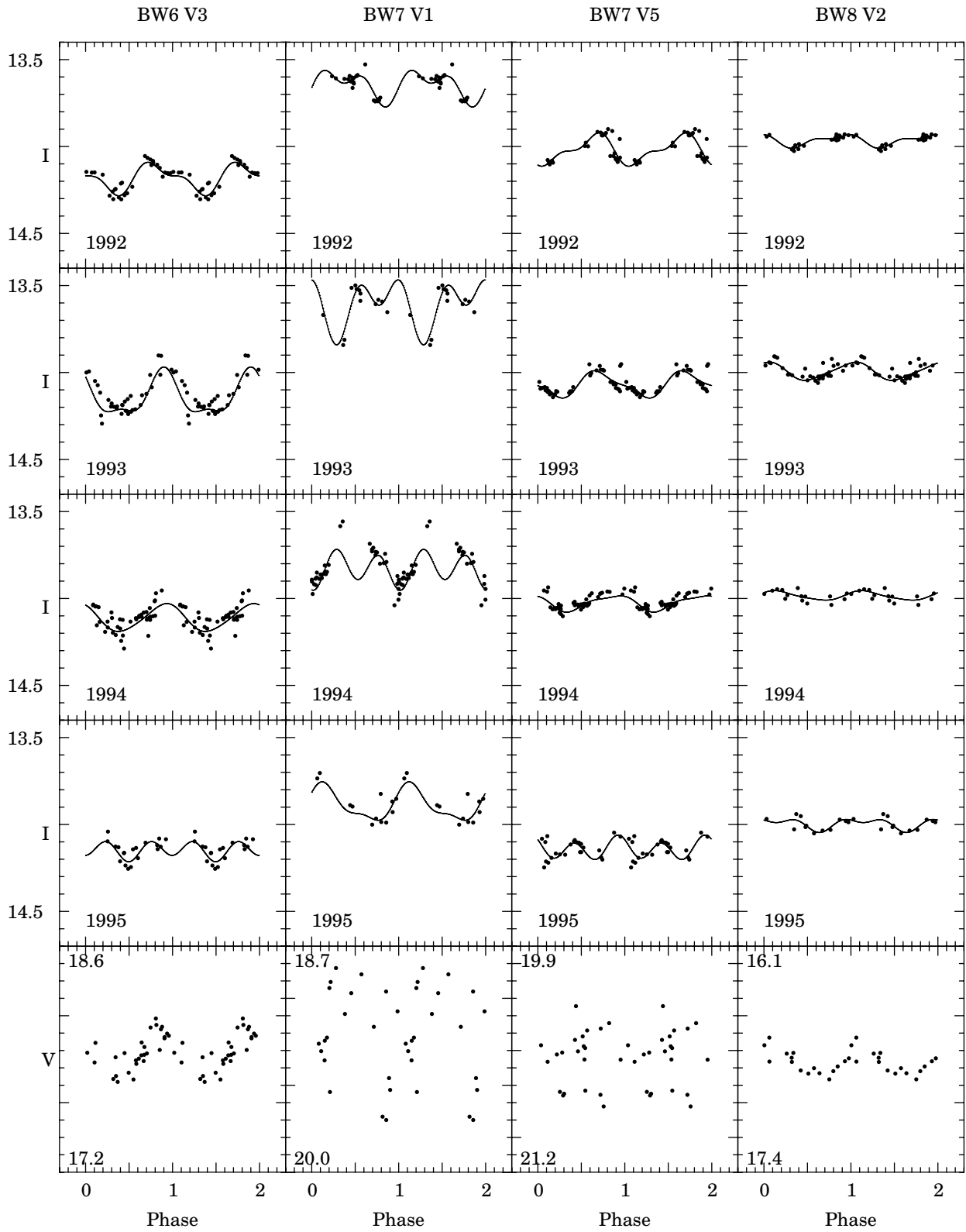
\end{document}